\documentclass[12pt]{iopart}

\usepackage{iopams}
\usepackage{setstack}

\usepackage[dvips]{graphicx}

\begin{document}

\title[Resonance production in heavy-ion collisions at STAR]
{Resonance production in heavy-ion collisions at STAR}

\author{Christina Markert \dag\ for the STAR collaboration
\footnote[2]{For the full author list and acknowledgements see
Appendix "Collaborations" in this volume.}}

\address{\dag\ Department of Physics, University of Texas at Austin, Austin, Texas 78712, USA}

\begin{abstract}
Hadronic resonances are sensitive to the properties of a hot and
dense medium created in a heavy ion collisions. During the hadronic
phase, after hadronization of quark and gluons into hadrons,
resonances are useful to determine the lifetime between chemical and
thermal freeze-out, under the assumption that the re-scattering of
the decay particles and the probability of regeneration of
resonances from hadrons depends on the system properties and the
resonance lifetime. The system size and energy dependence of
resonance spectra and yields will be shown and discussed in the
context of the lifetime and size of the hadronic phase. Elliptic
flow measurement will extend the sensitivity of resonance yields to
the partonic state through additional information on constituent
quark scaling. We also explore a possible new technique to extract
signals from the early, potentially chirally symmetric, stage
through the selection of resonances from jets.
\end{abstract}

\section{Introduction}
Due to the short lifetime of resonances (few fm/c), interactions
with the partonic and hadronic medium created in a heavy ion
reaction, can modify their properties. Depending on the medium
conditions; energy density, temperature and the degrees of freedom,
various modifications to resonance properties, such as mass, width
(lifetime), yields and spectra, are expected. The so-called bulk
matter of low momentum resonances, i.e. resonances with a transverse
momentum of p$_{\rm T} <$ 2~GeV/c, are dominated by the extended
dense hadronic medium due to an enhanced resonance regeneration
probability. Furthermore interactions of resonance decay hadrons
with the hadronic medium are present, which leads to a reduction of
the resonance yield. The reconstructed resonance yields, obtained
through invariant mass reconstruction based on hadronic decays, are
sensitive to the lifetime, density and temperature of the hadronic
medium and to the decay lifetimes of the resonances and their
regeneration cross sections. The measured ratio of the resonance to
non-resonance yields \cite{resostar} at $\sqrt{s_{\rm NN}} = $ 200
GeV in Au+Au collisions can be described with a microscopic
transport model (UrQMD) which assumes a duration between chemical
and kinetic freeze-out of $\Delta \tau = 10 \pm 3 $~fm/c
\cite{urqmd}. Alternatively, suppression of the $\Lambda$(1520) and
K(892) yields combined with a thermal model with an additional
re-scattering phase \cite{tor01,raf01,raf02,mar02}, suggests a
hadronic lifetime of $\Delta \tau
> 4$~fm/c \cite{resostar}. Together with the pion HBT lifetime
measurement ($\Delta\tau=5-12$~fm/c) ~\cite{nig05}, which determines
the time from the beginning of the collision to kinetic freeze-out,
a partonic lifetime can be extracted under the assumption that the
chemical freeze-out occurs at hadronization \cite{resostar}. The
further investigation of the lifetime and medium size dependencies
of the hadronic phase on the hadronic resonances is presented in
Section~\ref{chaptersize} and \ref{chapterenergy} using smaller
system size (Cu+Cu collisions) and a lower Au+Au collision energy
($\sqrt{s_{\rm NN}} = $ 62 GeV).

Another observable which is sensitive to the degrees of freedom in
the initial partonic phase, but which might also exhibit sensitivity
to contributions from regenerated resonances in the hadronic phase,
is the elliptic flow. The latter is the momentum anisotropy of
emitted particles due to the spatial azimuthal asymmetry of the
initial state under the assumption of maximum partonic interactions.
The so-called 'constituent quark scaling' of the elliptic flow for
resonances will increase if a resonance is re-combined from hadrons
rather than from quarks \cite{nonaka}. Results will be shown in
Section~\ref{chapterflow}.

In order to study the partonic and early hadronic medium with
resonances where chiral symmetry is possibly restored, the
resonances need to be unaffected by the late hadronic medium where
regeneration of resonances dilutes the signatures. Even the
reconstruction of leptonic decays might be affected by the signal
from the hadronically regenerated resonances. Since we know from
UrQMD calculations that the regeneration is predominant in the low
momentum region (p$_{\rm T} <$ 2~GeV/c) \cite{urqmd}, the study of
chiral symmetry restoration from leptonic decays might be only
suitable in the high momentum region of p$_{\rm T} >$ 2~GeV/c. The
same argument is valid for the hadronic decays. High momentum
resonances are produced early in the reaction and less affected by
the re-scattering and regeneration in the hadronic phase because the
system moves with a much smaller collective velocity and thus the
high momentum particles can escape, in particular if the hard
scattering which produced the resonance occurs near the surface.
Thus resonances from the same side jet in a triggered di-jet event
should be less suppressed since they most likely traveled through
less of the bulk medium. Therefore to investigate medium
modification of resonances in the medium resonances from the away
side, which have the longer path length in the medium need to be
selected. But their momenta should be large enough to be unaffected
by re-scattering and regeneration in the subsequently produced
hadronic medium. This idea requires the resonances to be formed
earlier than the bulk hadrons, such that hadronic resonances can
traverse the partonic medium. Formation time arguments will be
discussed in a later section. The analysis is based on selection of
resonances from the away-side of a di-jet using a high p$_{\rm T}$
hadron trigger to select the jet axis. Early feasibility studies
will be shown in Section~\ref{chapterjet}.

\section{System size dependence}
\label{chaptersize}

\begin{figure}[h!]
\centering
\includegraphics[width=0.46\textwidth]{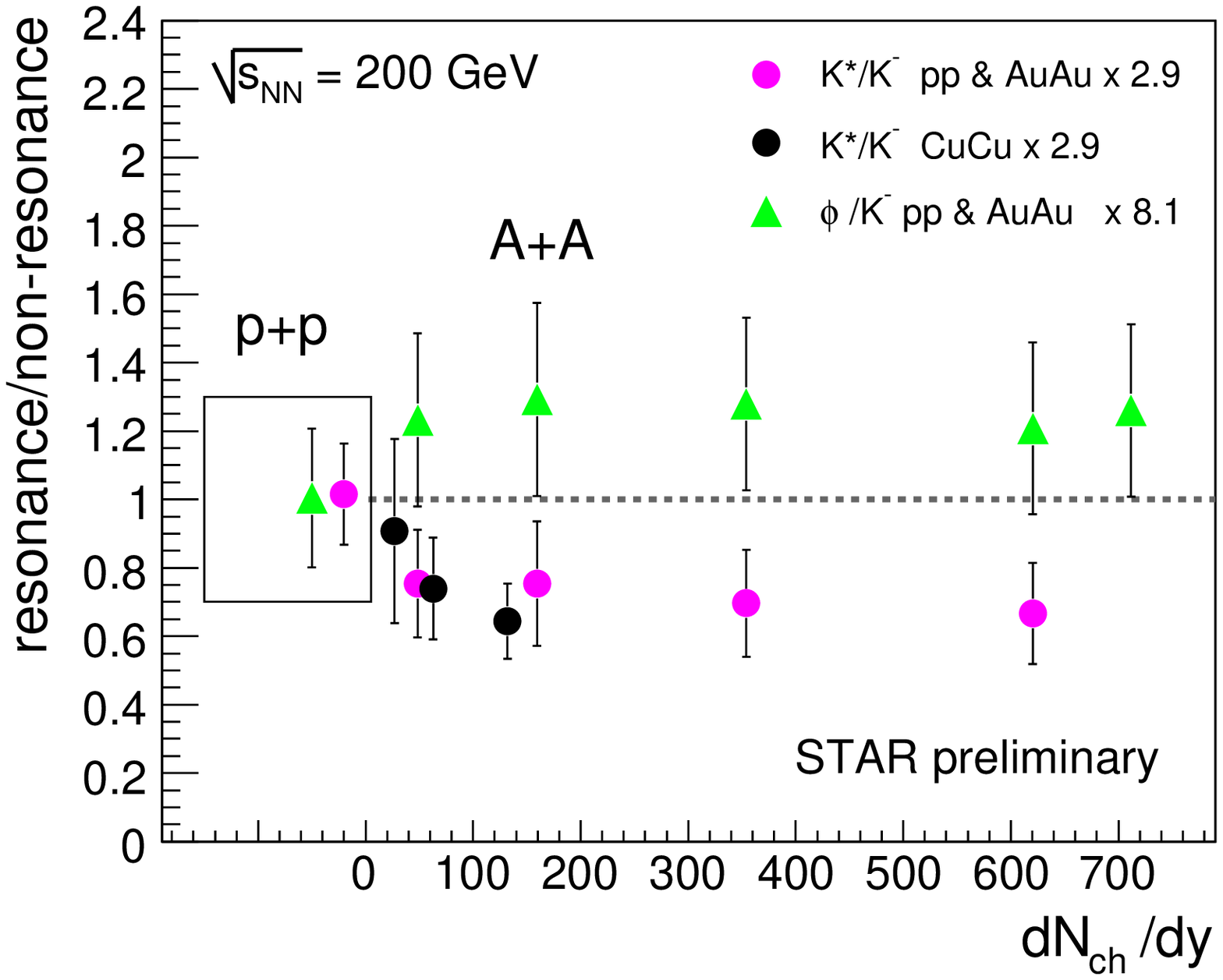}
\includegraphics[width=0.38\textwidth]{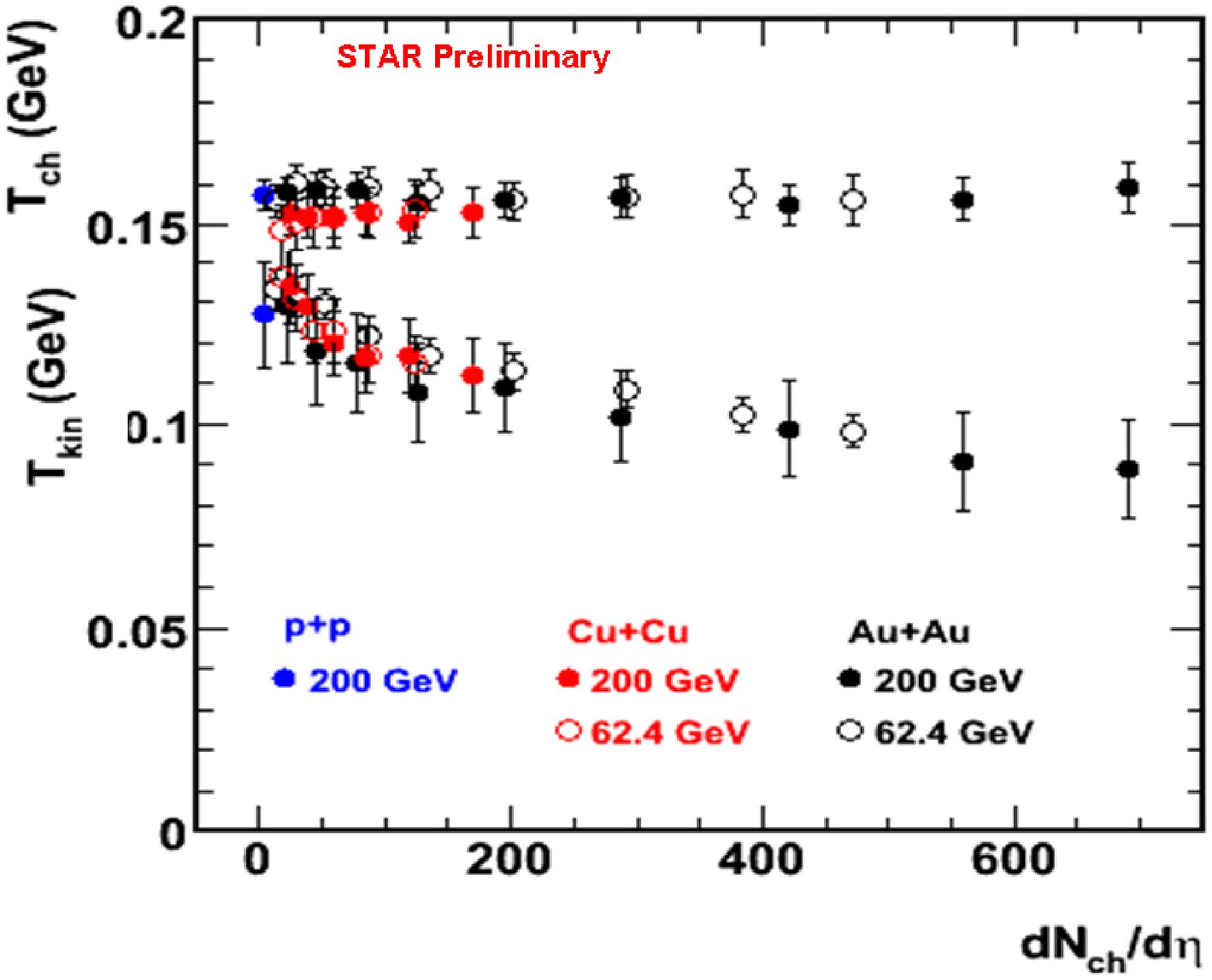}

\caption{Left: Ratio of $\rm K ^{*}/K^{-}$ and $\phi/K^{-}$ versus
dN$_{ch}$/d$\eta$ of Au+Au and Cu+Cu collisions at $\sqrt{s_{\rm
NN}} = $ 200 GeV. The ratios are normalized to unity in p+p
collisions. The Cu+Cu data only include statistical errors. Right:
Chemical and kinetic freeze-out temperatures versus
dN$_{ch}$/d$\eta$  \cite{aneta}.}
\vspace{-1cm}
 \label{cucu200}
\end{figure}

The lifetime of the hadronic phase depends on the system size and
the expansion velocity. Elementary p+p collisions exhibit a small
system size and a nearly zero lifespan of the hadronic phase. We
studied the ratios of resonances to stable particles with respect to
centrality defined by the multiplicity of produced charged
particles. We measured a suppression of $\rm K ^{*}/K^{-}$
\cite{zhang} and $\Lambda ^{*}/\Lambda$ \cite{resostar} in central
Au+Au compared to minimum bias p+p collisions. Due to large errors
the onset of the suppression from minimum bias p+p to peripheral
Au+Au collisions could not be assigned to a narrow multiplicity
range. With the new $\rm K ^{*}/K^{-}$ \cite{sadhana,aneta} from
Cu+Cu collisions we are able to better differentiate the low
multiplicity region due to the smaller system size of the Cu nuclei.
Fig.~\ref{cucu200} shows the ratio of $\rm K ^{*}/K^{-}$ versus the
charged particle multiplicity for Au+Au and Cu+Cu collisions at
$\sqrt{s_{\rm NN}}=$ 200 GeV. The agreement in the data shows that
the suppression of resonances scales with charged particle
multiplicity. Furthermore the onset of the suppression seems to be
occurring at a very low multiplicity and the maximum suppression is
reached already at dN$_{ch}$/d$\eta$ = 100 and remains constant out
to the most cental Au+Au collision (dN$_{ch}$/d$\eta$ = 700). This
implies that the hadronic lifetime between the peripheral and
central collisions remains constant. This trend is similar for the
temperature separation between chemical and thermal freeze-out shown
in figure~\ref{cucu200} (left).

\section{Energy dependence}
\label{chapterenergy}

\begin{figure}[h!]
\centering
\includegraphics[width=0.48\textwidth]{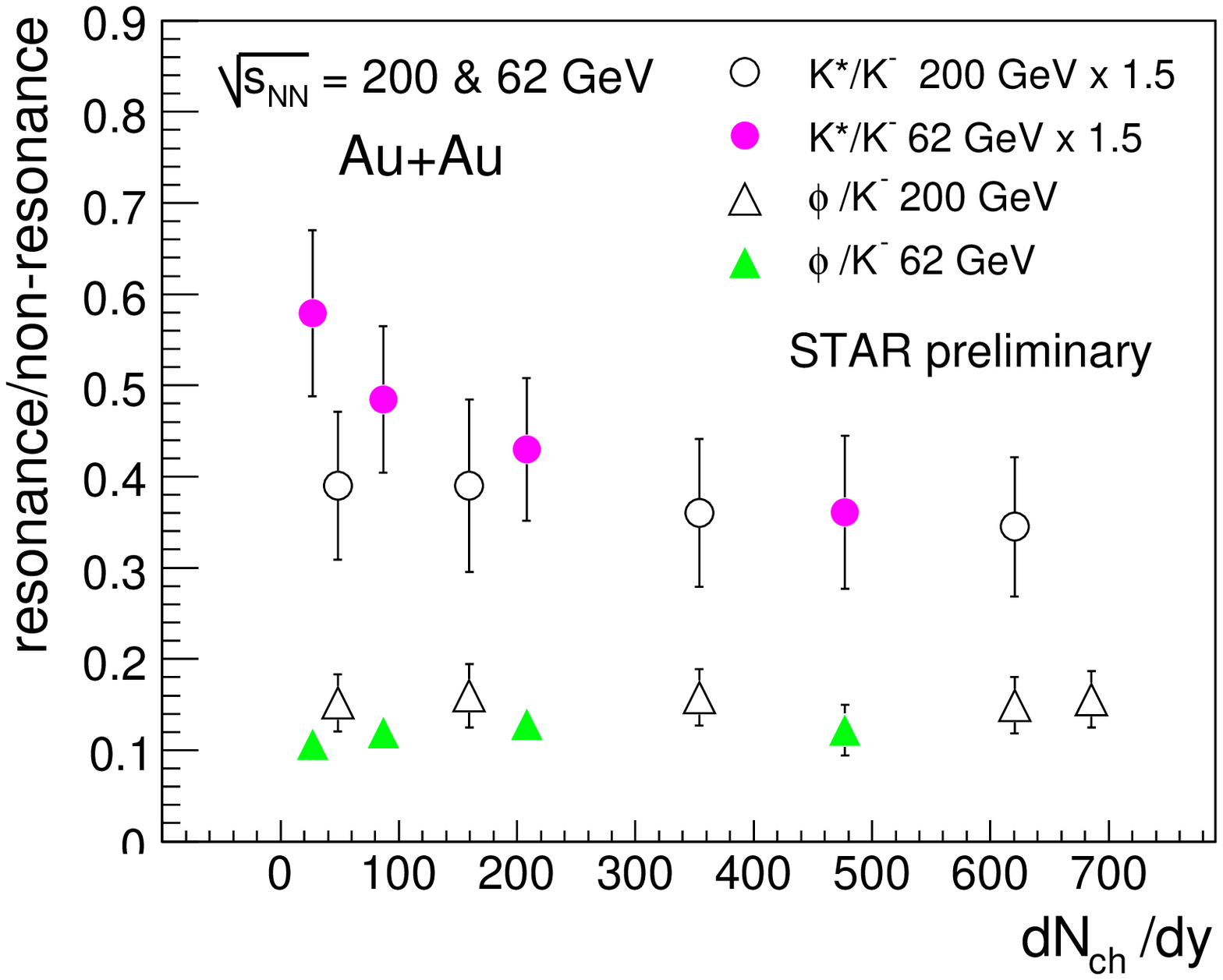}
\includegraphics[width=0.48\textwidth]{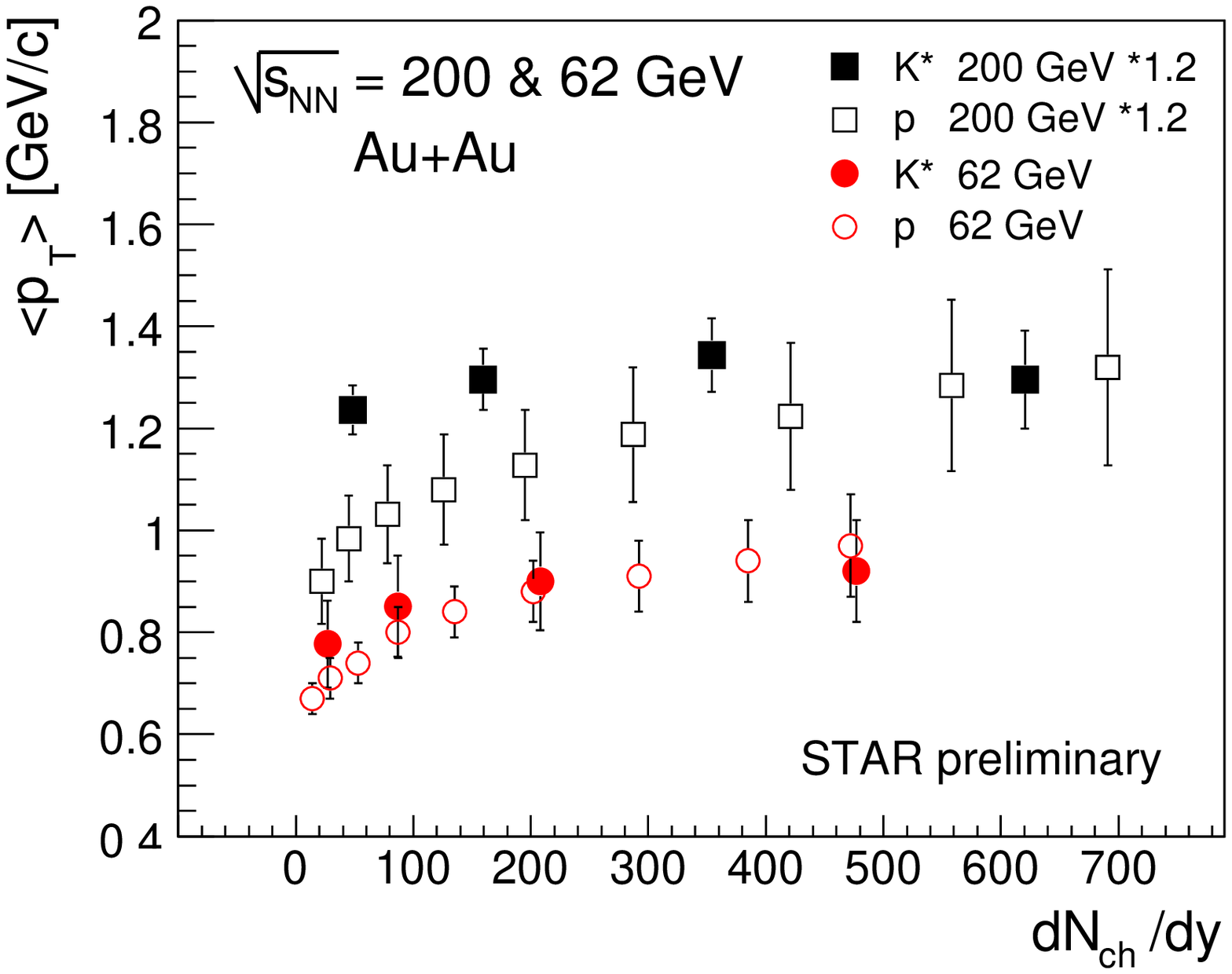}

\caption{Left: Ratio of $\rm K ^{*}/K^{-}$ and $\phi/K^{-}$ versus
dN$_{ch}$/d$\eta$ of Au+Au collisions at $\sqrt{s_{\rm NN}} = $ 200
GeV and $\sqrt{s_{\rm NN}} = $ 62 GeV. Right: The $\langle p_{\rm T}
\rangle$ versus dN$_{ch}$/d$\eta$ of $\rm K ^{*}$ and a non-resonant
particle of similar mass (proton) in Au+Au collisions at
$\sqrt{s_{\rm NN}} = $ 200 GeV and $\sqrt{s_{\rm NN}} = $ 62 GeV.}
\label{auau62}
\end{figure}

Fig.~\ref{auau62} (left) shows the ratio of $\rm K ^{*}/K^{-}$ and
$\phi/K^{-}$ versus the charged particle multiplicity of Au+Au
collisions at two different energies, $\sqrt{s_{\rm NN}} = $ 200 GeV
and $\sqrt{s_{\rm NN}} = $ 62 GeV. The $\rm K ^{*}/K^{-}$ data
indicate a smaller suppression at lower energies in the most
peripheral collisions. In central collisions the suppression of the
$\rm K ^{*}/K^{-}$ is the same for both collision energies. This
would indicate that the hadronic lifetime or the re-scattering cross
section increase, as a function of centrality, is slower in lower
energy collisions. The mean transverse momentum shown in
Fig.~\ref{auau62} (right) confirms the larger re-scattering
contribution, which increases the mean transverse momentum in
peripheral collisions at the higher incident energy.

\begin{figure}[h!]
\centering
\vspace{-0.5cm}
\includegraphics[width=0.46\textwidth]{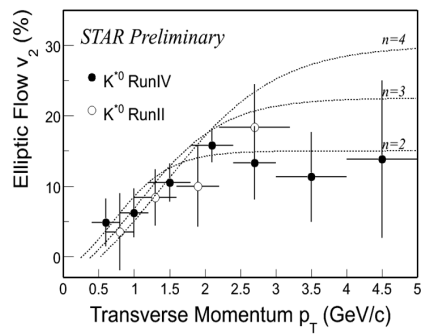}
\vspace{-1.0cm} \caption{Elliptic flow of $\rm K ^{*}$ versus the
transverse momentum. The lines show the predictions for a
constituent quark scaling of 2, 3 and 4. \cite{zhang,nonaka}}
\label{flow}
\end{figure}

\section{Elliptic flow}
\label{chapterflow}

Based on constituent quark scaling considerations, the elliptic flow
of directly produced $\rm K ^{*}$ from partons should scale with two
(two constituent quarks), while the flow of re-combined $\rm K
^{*}$, which are generated from a two meson interaction, should
scale with four (four constituent quarks) \cite{nonaka}. According
to Nonaka's calculations the elliptic flow of $\rm K ^{*}$ will
increase by about 15\% in the transverse momentum region of 3-5
GeV/c if both contributions are mixed according to a recombination
model. Presently the error in our $\rm K ^{*}$ flow determination,
shown in Figure~\ref{flow} \cite{zhang}, is too large and the
measurement is not sufficiently sensitive to observe a possible
increase of the elliptic flow. The error will improve with the run 7
data.

\section{Resonances from Jets}
\label{chapterjet}

\begin{figure}[h!]
\centering
\includegraphics[width=0.65\textwidth]{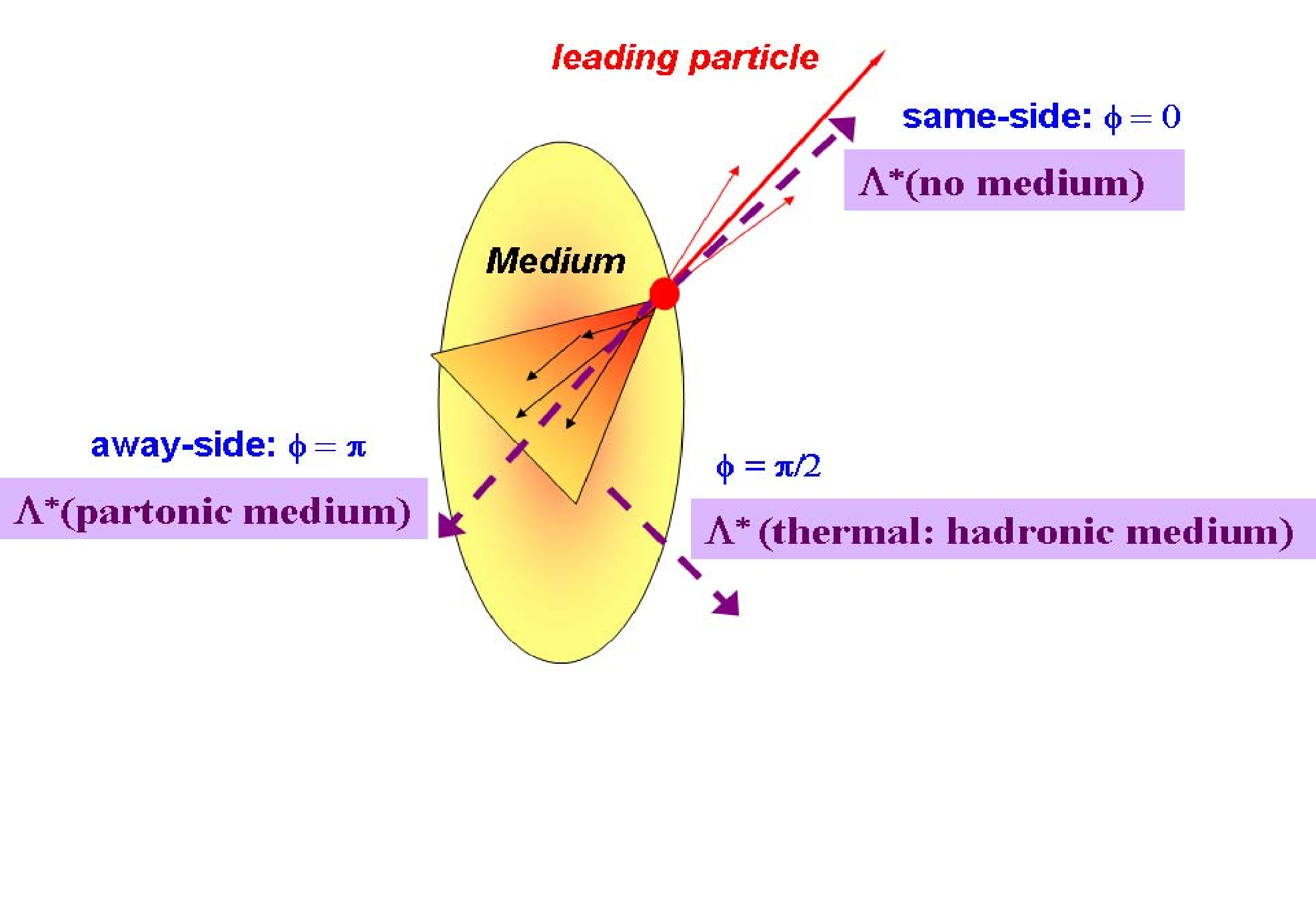}
 \caption{Sketch of jet fragmentation into
resonances ($\Lambda$*, $\phi$(1020),...) in the medium created in a
heavy-ion collision. Same-side correlations of resonances are not
affected by the medium, whereas the away-side high p$_{\rm T}$
resonance might be affected by the early (chiral symmetry restored)
medium. Thermal resonances, which are affected by the late hadronic
medium are at $\pi$/2 with respect to the trigger particle.}
 \label{jetresosketch}
\end{figure}

The isolation of resonances from an away-side jet which passes
through the partonic medium might be a suitable measurement to
investigate the effect of the partonic or early hadronic medium on
resonances with respect to mass shifts and width broadenings
\cite{marjet}. High momentum resonances from the away-side jet are
identified via the angle with respect to the jet axis or leading
particle (see Figure~\ref{jetresosketch}). A high transverse
momentum resonance in the away-side jet cone is likely to be
produced early, which, depending on its formation time, can interact
with the early partonic medium but leaves the medium quickly enough
to not exhibit any interaction with the later hadronic phase. The
formation time of a resonance in the string fragmentation process
depends on the momentum fraction $z$ carried by the resonance. In
addition there is a parton and resonance mass dependence which leads
to shorter formation times for heavy resonances. A quantum
mechanical treatment of heavy meson formation in heavy-ion
collisions is shown in \cite{vitev}. The authors demonstrate that
the probability of high momentum heavy hadron (or resonance)
formation in the partonic medium is finite. An alternate approach,
based on string fragmentation \cite{falter}, arrives at a similar
conclusion for heavy mass, light quark objects. Quantitative studies
of resonance properties such as yield, mass, width, and branching
ratio as a function of resonance momentum, emission angle, jet
energy, and jet tag, might therefore directly address the question
of chiral symmetry restoration.

\begin{figure}[h!]
\centering
\includegraphics[width=0.50\textwidth]{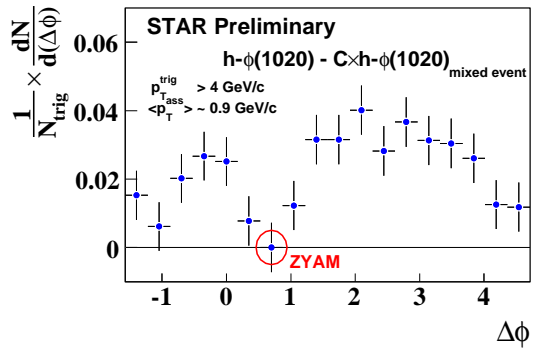}
\vspace{-0.6cm} \caption{STAR, angular correlation of
hadron-$\phi$(1020) resonance. Hadron trigger p$_{\rm T}$ $>$
4~GeV/c and associated $\phi$(1020) $\langle p_{\rm T} \rangle \sim
0.9$ GeV/c \cite{marjet}.}
\label{jet}
\vspace{-0.5cm}
\end{figure}

A first attempt of this correlation analysis was done using the
$\phi(1020)$ meson and a charged leading particle \cite{marjet}. The
masses and widths of the $\phi$(1020) signals for the different
angular selections are in agreement with the PDG value. Due to the
low momentum of the associated $\phi$(1020) particle most of them
are from the late hadronic medium and therefore we are not sensitive
to mass shifts or width broadenings. Alternatively $\phi$(1020) are
selected via invariant mass cut and correlated with a high momentum
trigger hadron. Figure~\ref{jet} shows the raw hadron-$\phi$(1020)
correlation after normalization and subtraction of the mixed-event
background with the lowest point at zero. This preliminary result is
not corrected for elliptic flow (v$_{2}$) contribution and has no
systematic error estimation. However the trend of a larger resonance
production on the away-side of the $\Delta \phi$ correlation
compared to the same side is present in the angle dependent mass
distribution. This effect might be due to energy conservation
(trigger bias), however. More statistics in the present analysis as
well as production runs with the new Time of Flight upgrade detector
in the future will help us to identify the resonance decay daughters
out to higher momenta and therefore reduce the combinatorial
background up to factor of 10.

\section{Conclusion}
Hadronic decays of resonances with different lifetimes are used to
extract information about the time evolution and temperature of the
expanding hadronic medium. Data from the smaller system (CuCu)
exhibit the same dependence of the resonance suppression versus the
charged particle production as measured in AuAu. The $\rm K
^{*}/K^{-}$ ratios from Cu+Cu collisions indicate that the hadronic
lifetime remains constant from 200 out to 700 produced charged
particles (per rapidity unit). The $\sqrt{s_{\rm NN}} = $ 62 GeV
data indicate a slower decrease of the $\rm K ^{*}/K^{-}$ ratio and
a slower increase of mean p$_{\rm T}$ versus the centrality compared
to the $\sqrt{s_{\rm NN}} = $ 200 GeV data. These results can be
interpreted as a slower increase of the hadronic lifetime at lower
collision energies. Resonances from jets are being investigated as a
tool to access early-produced resonances, which are unaffected by
the hadronic medium, in order to study chiral symmetry restoration.

\end{document}